\documentclass[twocolumn,aps,prl,superscriptaddress,floatfix]{revtex4-2}

\usepackage{amsmath,amssymb,amsfonts}
\usepackage{graphicx}
\usepackage{bm}
\usepackage{xcolor}
\usepackage{hyperref}
\usepackage{dsfont}
\usepackage{bbm}
\usepackage{braket}
\usepackage{tikz}
\usetikzlibrary{decorations.markings,arrows.meta}

\newcommand{\Tr}{\mathrm{Tr}}

\newcommand{\Wcp}{W_{\rm CP}}
\newcommand{\Qcp}{Q_{\rm CP}}
\newcommand{\expect}[1]{\langle #1 \rangle}

\newcommand{\mtau}{m_\tau}
\newcommand{\hllhc}{\text{HL-LHC}}

\begin{document}

\title{Quantum Magic Reveals CP Phases Invisible to Entanglement in Spin-0 Decays}

\author{Nicol\'as Viaux\,$^{\ast}$}
\affiliation{Departamento de F\'isica, Universidad T\'ecnica Federico Santa Mar\'ia, Casilla 110-V, Valpara\'iso, Chile}
\affiliation{Millennium Institute for Subatomic Physics at High Energy Frontier (SAPHIR), Santiago, Chile}

\author{Ariel Norambuena\,$^{\dagger}$}
\affiliation{Departamento de F\'isica, Universidad T\'ecnica Federico Santa Mar\'ia, Casilla 110-V, Valpara\'iso, Chile}

\author{Pedro Orellana\,$^{\ddagger}$}
\affiliation{Departamento de F\'isica, Universidad T\'ecnica Federico Santa Mar\'ia, Casilla 110-V, Valpara\'iso, Chile}

\date{\today}

\begin{abstract}
All standard scalar quantum-information measures---concurrence, negativity, entanglement entropy, the optimized CHSH bound, and quantum Fisher information---are CP-blind in ideal \\ spin-0 $\to f\bar f$ decays because the two-qubit spin state is maximally entangled for every CP angle. We show that stabilizer magic, fixed in the physical Pauli frame of spin analysis, escapes this blind spot: the stabilizer R\'enyi entropy admits an exact closed form, vanishing at CP-definite and Clifford phases and peaking at maximal non-Clifford mixing. Two experimentally accessible, magic-inspired CP witnesses follow; the linear amplitude is $14.3\times$ more efficient than its quartic counterpart and reaches discovery-level sensitivity at the HL-LHC for $H\to\tau^+\tau^-$.
\end{abstract}

\maketitle

\vspace{-2ex}
\begingroup\scriptsize\setlength{\baselineskip}{8pt}
\noindent$^{\ast}$\href{mailto:nicolas.viaux@usm.cl}{nicolas.viaux@usm.cl}\\
\noindent$^{\dagger}$\href{mailto:ariel.norambuena@usm.cl}{ariel.norambuena@usm.cl}\\
\noindent$^{\ddagger}$\href{mailto:pedro.orellana@usm.cl}{pedro.orellana@usm.cl}\par
\endgroup
\vspace{0.3ex}

\textit{Introduction.---} Quantum information science is emerging as a new lens to interrogate fundamental physics at the highest accessible energies~\cite{QIHEP1,QIHEP2}. The first observations of entanglement in $t\bar t$ at ATLAS~\cite{ATLAS_ttbar} and CMS~\cite{CMS_ttbar}, together with the recent measurement of quantum magic in $t\bar t$ by CMS~\cite{CMS_magic}, have established colliders as platforms for testing the foundations of quantum mechanics far beyond conventional laboratory scales.

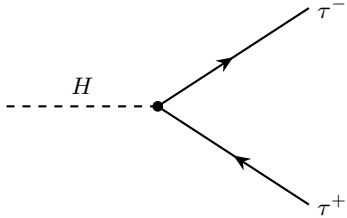
\begin{figure}[htb]
  \centering
  \begin{tikzpicture}[
    every node/.style={font=\small},
    farr/.style={postaction={decorate},decoration={markings,
                 mark=at position 0.5 with
                 {\arrow{Stealth[length=2.2mm,width=1.8mm]}}}}
  ]
    \draw[dashed,thick] (-2.0,0) -- (0,0);
    \node[above] at (-1.0,0.05) {$H$};
    \filldraw (0,0) circle (1.8pt);
    \draw[farr,thick] (0,0) -- (2.0,1.3);
    \node[right] at (2.0,1.3) {$\tau^-$};
    \draw[farr,thick] (2.0,-1.3) -- (0,0);
    \node[right] at (2.0,-1.3) {$\tau^+$};
  \end{tikzpicture}
  \caption{Tree-level $H\to\tau^+\tau^-$ via the CP-mixed Yukawa of Eq.~(\ref{eq:lagrangian}); same topology for any spin-0 $\to f\bar f$ decay.}
  \label{fig:feynman}
\end{figure}

The CP structure of the Higgs Yukawa couplings remains a key unresolved issue at the LHC. The most general Lorentz-invariant interaction governing $H\to\tau^+\tau^-$ is
\begin{equation}
  \mathcal{L} \supset -\frac{\mtau}{v}\,H\,\bar{\tau}\!\left(\cos\alpha
  + i\sin\alpha\,\gamma_5\right)\!\tau,
  \label{eq:lagrangian}
\end{equation}
with $\alpha=0$ the SM scalar and $\alpha=\pi/2$ a pure pseudoscalar; the corresponding tree-level topology, illustrated in Fig.~\ref{fig:feynman}, applies to any spin-0 $\to f\bar f$ decay with CP-mixed Yukawa. ATLAS and CMS bounds give $\phi_\tau = 9^\circ\pm16^\circ$ and $-1^\circ\pm19^\circ$~\cite{ATLAS_Htautau,CMS_Htautau}, with HL-LHC expected to push sensitivity further~\cite{Barr_review}.

Entanglement in $H\to\tau^+\tau^-$ is maximal for all $\alpha$~\cite{Altakach_Htautau,Fabbrichesi_Htautau}; magic in $t\bar t$~\cite{CMS_magic,White_magic,QED-magic} adds a further QI observable to the broader QI-at-colliders program~\cite{Afik_ttbar,Barr_review,Altakach_polarized,Ai_QFI}. Whether any QI \emph{resource} diagnostic actually probes the Yukawa CP angle, however, has remained open: spectrum-based monotones (concurrence, negativity, CHSH) and quantum Fisher information are $\alpha$-independent because the state is maximally entangled. Classical spin-correlation measurements---tomography of the $C_{ij}$ matrix and the acoplanarity angle---do probe CP through kinematic angular dependences, but they carry no quantum-information content. The CP angle survives in the \emph{orientation} of the spin-correlation tensor, which is not itself a resource-theoretic QI scalar.

In this Letter we show that stabilizer magic probes precisely that orientation. The stabilizer R\'enyi entropy (SRE) depends on fourth-order Pauli moments and so resolves whether the CP-induced spin phase is Clifford or non-Clifford in the physical spin-analysis frame---an algebraic property invisible to any spectrum-based functional. We derive the SRE in closed form, identify two experimentally accessible magic-inspired CP witnesses, and establish their reach at the HL-LHC.

\textit{The $H\to\tau^+\tau^-$ two-qubit state.---} In the rest frame of the decaying particle, with $\tau^+$ momentum as the quantization axis $\hat z$, Eq.~(\ref{eq:lagrangian}) prepares~\cite{Altakach_Htautau,Fabbrichesi_Htautau}
\begin{equation}
\ket{\psi(\alpha)}
=
\frac{1}{\sqrt{2}}\left(\ket{\uparrow\downarrow}+ e^{2 i \alpha}\ket{\downarrow\uparrow}\right),
\label{eq:state}
\end{equation}
maximally entangled for every $\alpha\in\mathbb{R}$ and obtained from the Bell state $\ket{\Psi^+}=(\ket{\uparrow\downarrow}+\ket{\downarrow\uparrow})/\sqrt 2$ by adding a local $\hat z$ rotation. This already anticipates why all local-unitary-invariant entanglement measures are independent of $\alpha$: $\alpha$ is gauge under local unitaries acting on the reduced states. The local Bloch vectors vanish, so all spin information lies in $C_{ij}(\alpha)\equiv\Tr[\rho(\alpha)\sigma_i\otimes\sigma_j]$, with $\rho(\alpha)=\ket{\psi(\alpha)}\!\bra{\psi(\alpha)}$ and
\begin{equation}
C(\alpha)=
\begin{pmatrix}
\cos 2\alpha & -\sin 2\alpha & 0 \\
\sin 2\alpha & \cos 2\alpha  & 0 \\
0            & 0             & -1
\end{pmatrix},
\label{eq:correlation_matrix}
\end{equation}
an improper orthogonal matrix with $\det C(\alpha)=-1$ regardless of $\alpha$. The CP angle is therefore a rotation of the transverse $xy$ block that preserves its singular values; the off-diagonal entries $C_{xy}=-\sin 2\alpha$ and $C_{yx}=\sin 2\alpha$ carry the explicit CP-sensitive content and break the symmetry between $C_{ij}$ and $C_{ji}$.

\begin{figure}[htb]
  \centering
  \includegraphics[width=\columnwidth]{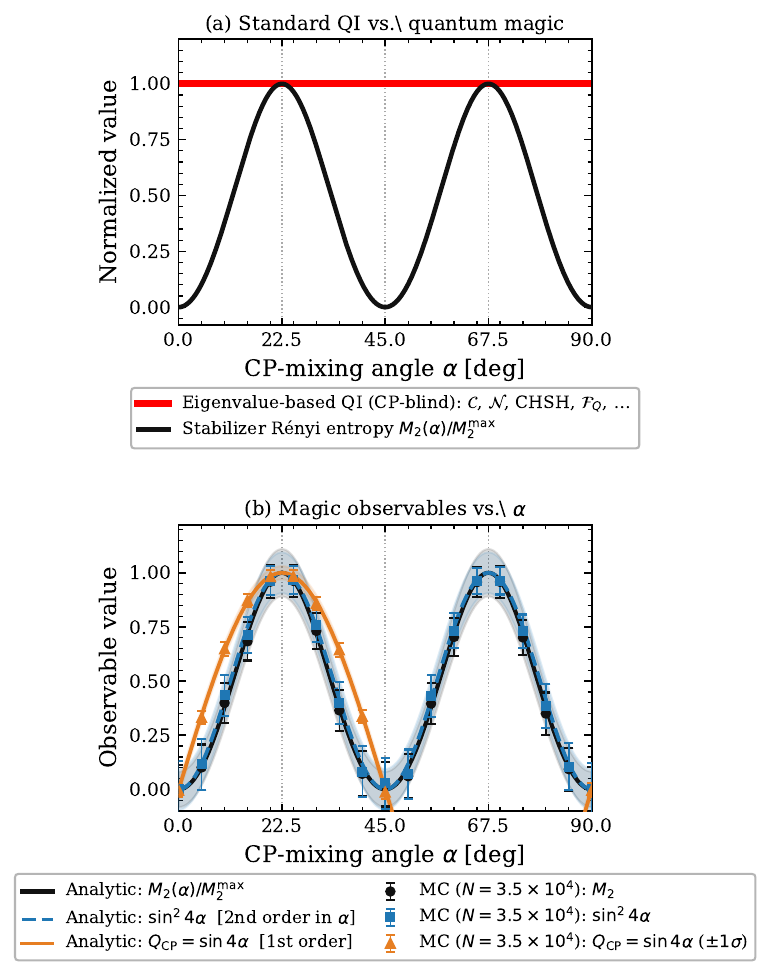}
  \caption{(a) Standard QI measures (concurrence $\mathcal C$, negativity $\mathcal N$, von Neumann entropy $S$, CHSH bound, quantum Fisher information $\mathcal F_Q$), normalized to their maxima, collapse onto the same horizontal line independent of $\alpha$ (red band); the stabilizer R\'enyi entropy $M_2/M_2^{\max}$ (black) is the unique exception. (b) Magic observables from $10^6$ simulated $pp\to H\to\tau^+\tau^-$ events: $M_2/M_2^{\max}$ (black circles), $4-\Wcp=\sin^2 4\alpha$ (blue squares, quadratic in $\alpha$), $\Qcp=\sin 4\alpha$ (orange triangles, linear and signed). Shaded $\pm1\sigma$ bands at HL-LHC yield $N=3.5\times10^4$ $\pi\pi$ events~\cite{Fabbrichesi_FCC}. Dotted verticals: $\alpha=0,\pi/2$ (CP-definite), $\pi/4$ (Clifford), $\pi/8$ (max magic).}
  \label{fig:magic_vs_alpha}
\end{figure}

\textit{CP blind spot of standard QI observables.---} Maximal entanglement gives reduced density matrices $\rho_A=\rho_B=\openone/2$, with spectrum $\{1/2,1/2\}$ independent of $\alpha$. All spectrum-based QI indicators are therefore constant: the concurrence is $\mathcal{C}=1$, the negativity is $\mathcal{N}=1/2$, the von Neumann entropy is $S=1$~bit, the optimized CHSH violation is $S_{\max}=2\sqrt{2}$, and the quantum Fisher information of the pure-state family with respect to $\alpha$ is $\mathcal{F}_Q=4$. Each sits at its own numerical value, but once normalized to its respective maximum they all collapse to the same horizontal line at unity, shown as the red band in Fig.~\ref{fig:magic_vs_alpha}(a).

The CP parameter is not encoded in the \emph{amount} of entanglement but in the \emph{orientation} of the two-spin correlation pattern: any scalar monotone built solely from the spectrum of the density matrix $\rho$ or its reduced density states loses this information. The CP-sensitive content lives in functionals that probe the ordered structure of $C(\alpha)$ in a fixed physical frame.

\textit{Quantum magic and exact closed form.---} The stabilizer R\'enyi entropy (SRE) of order two~\cite{Leone_SRE,Oliviero_SRE} for an $n$-qubit pure state is
\begin{equation}
M_2(\psi)= -\log_2\!\left(\frac{1}{4^n}\sum_{P\in\mathcal{P}_n}\langle P\rangle^4\right)-n,
\label{eq:M2_def}
\end{equation}
with $\mathcal{P}_n=\{I,X,Y,Z\}^{\otimes n}$ the phase-free Hermitian Pauli strings. $M_2=0$ characterizes stabilizer states, which together with Clifford operations and Pauli measurements are classically simulable via the Gottesman--Knill theorem~\cite{Gottesman_stabilizer}; $M_2>0$ certifies nonstabilizerness, or quantum magic~\cite{Campbell_magic}. Unlike spectrum-based indicators, $M_2$ probes fourth-order Pauli moments---the ordered structure of the correlation tensor in the physical spin-analysis frame.

From Eq.~(\ref{eq:state}) with $n=2$, we note that all one-body expectations vanish; only the identity and the five two-body correlations of $C(\alpha)$ contribute. Using $\cos^4 x+\sin^4 x=1-\tfrac{1}{2}\sin^2 2x$, we obtain
\begin{equation}
\Xi(\alpha)\equiv\sum_{P\in\mathcal{P}_2}\langle P\rangle^4=4-\sin^2 4\alpha,
\label{eq:Xi}
\end{equation}
giving the exact closed form
\begin{equation}
M_2(\alpha)=\log_2\!\left(\frac{4}{4-\sin^2 4\alpha}\right).
\label{eq:M2_exact}
\end{equation}
$M_2(\alpha)\ge0$ vanishes precisely at $\alpha=k\pi/4$ ($k\in\mathbb{Z}$), the angles at which the relative phase $e^{2i\alpha}\in\{1,i,-1,-i\}$ is a Clifford phase, and reaches
\begin{equation}
M_2^{\max}=\log_2\!\left(\frac{4}{3}\right)\approx 0.415\;\text{bits}
\end{equation}
at $\alpha=\pi/8+k\pi/4$, where $e^{2i\alpha}=e^{i\pi/4}$ is the canonical T-gate phase, maximally non-Clifford within this family. The SRE is therefore not a monotone of the scalar-pseudoscalar admixture itself---it also vanishes at the CP-mixed Clifford angles $\alpha=\pi/4, 3\pi/4$, where the Yukawa coupling is still CP-mixed but the induced two-qubit phase is classically simulable. What $M_2$ quantifies is the non-Clifford content of the CP-induced spin phase; it is the exact CP-sensitive resource-theoretic discriminator complementary to standard QI [Fig.~\ref{fig:magic_vs_alpha}(b)].

\textit{Magic Witnesses $\Wcp$ and $\Qcp$.---} The quartic Pauli sum defines a CP-sensitive witness,
\begin{equation}
  \Wcp \equiv \sum_{P\in\mathcal{P}_2}\!\expect{P}^4 = 4-\sin^2 4\alpha \in [3,4],
  \label{eq:Wcp}
\end{equation}
with $M_2=\log_2(4/\Wcp)$. Using $C_{xx}=C_{yy}$ and $|C_{xy}|=|C_{yx}|$,
\begin{equation}
  \Wcp = 2 + 2\!\left(C_{xx}^4 + C_{xy}^4\right).
  \label{eq:Wcp_exp}
\end{equation}
Within the ideal pure-state spin family, $\Wcp<4$ is equivalent to nonstabilizerness and therefore to nonzero $M_2$. In realistic reconstructed samples, where acceptance, detector effects, backgrounds, and radiative corrections induce mixing, $\Wcp$ should be interpreted as a magic-inspired CP witness rather than a strict magic monotone, unless a mixed-state magic analysis is performed. Crucially, $\Wcp=4$ does \emph{not} imply CP conservation, since the Clifford angles $\alpha=m\pi/4$ also yield stabilizer states despite a CP-mixed Yukawa.

A complementary observable with better statistical reach is the signed magic-sensitive amplitude
\begin{equation}
  \Qcp \equiv \sin 4\alpha = -2\,C_{xy}\,C_{xx},
  \label{eq:Qcp}
\end{equation}
whose square fixes the SRE, $M_2(\alpha)=\log_2[4/(4-\Qcp^2)]$. Near the SM point, $\Qcp\simeq 4\alpha$ is linear in $\alpha$, whereas $4-\Wcp\simeq 16\alpha^2$ is quadratic. This first- vs.\ second-order signal structure produces the empirical Monte Carlo ratio $N_{5\sigma}[\Wcp]/N_{5\sigma}[\Qcp]=14.3$ quoted below.

\begin{figure*}[t]
  \centering
  \includegraphics[width=\textwidth]{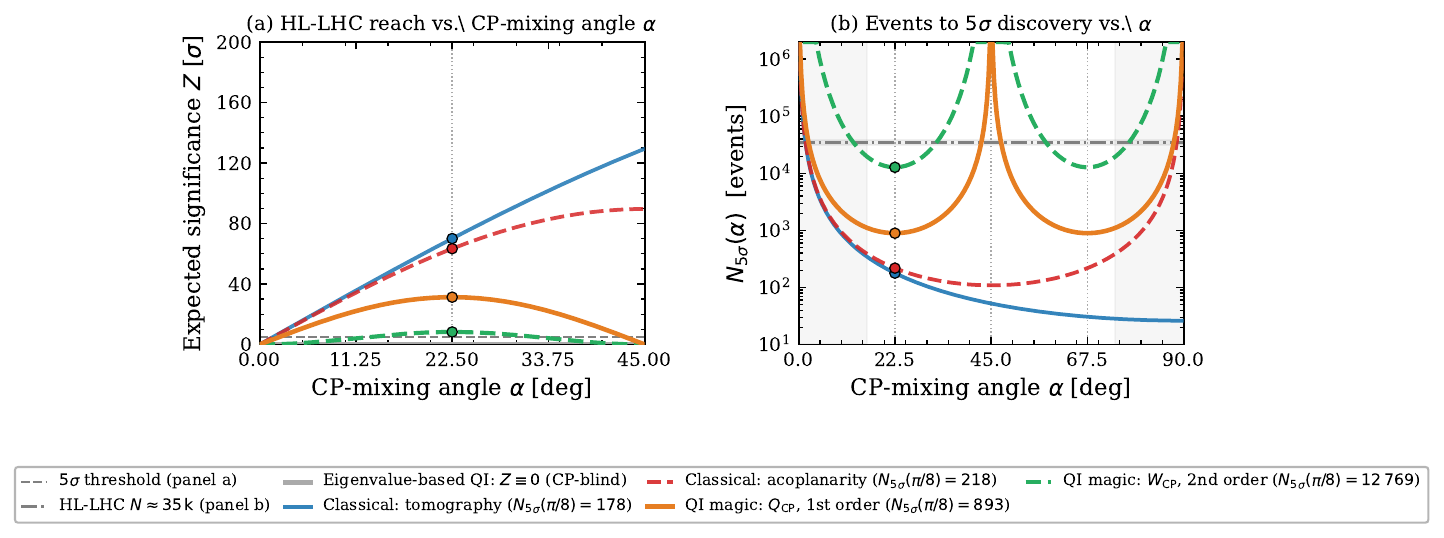}
  \caption{Three-way comparison at HL-LHC ($N\approx35{,}000$ events), idealized spin-correlation baseline (cf.~\cite{Altakach_Htautau} for full-simulation). (a)~Discovery significance $Z(\alpha)$: classical tomography (blue solid) and acoplanarity $\phi^*$ (red dashed), and magic witnesses $\Qcp$ (orange solid), $\Wcp$ (green dashed), are CP-sensitive; spectrum-based QI is CP-blind ($Z\equiv0$, gray line); filled circles mark $Z(\pi/8)$. (b)~Events for $5\sigma$ discovery $N_{5\sigma}(\alpha)$: all curves diverge at $\alpha=0,\pi/2$ (CP-conserving); $\Qcp$, $\Wcp$ additionally vanish at the Clifford angle $\alpha=\pi/4$ (stabilizer despite CP-mixed Yukawa). Dot-dashed line: HL-LHC yield; shaded wings: angles excluded at 68\% CL by ATLAS. Filled circles: $N_{5\sigma}(\pi/8)$.}
  \label{fig:sensitivity}
\end{figure*}

\textit{Statistical sensitivity at HL-LHC.---} Figure~\ref{fig:sensitivity} compares four CP observables at HL-LHC ($N_{\pi\pi}\approx 35{,}000$ events~\cite{Fabbrichesi_FCC}). For $\tau^\pm\to\pi^\pm\nu$ the pion direction in the $\tau$ rest frame is the maximally efficient spin analyzer. Two classical benchmarks---full spin tomography of $C_{ij}$~\cite{Altakach_Htautau} and the acoplanarity angle $\phi^*$~\cite{CMS_Htautau,ATLAS_Htautau}---probe CP through kinematic correlations but are not QI scalars. Standard spectrum-based QI ($\mathcal{C}$, CHSH, $\mathcal{F}_Q$) is CP-blind, $Z\equiv 0$. The magic witnesses $\Qcp$ and $\Wcp$, derived from the SRE, are QI-based CP probes; both peak at $\alpha=\pi/8$. Panel~(b) shows the complementary discovery frontier $N_{5\sigma}(\alpha)$: each curve diverges at the nodes of its signal function and reaches its minimum at maximum signal strength.

\textit{Monte Carlo and measurement.---} Pion polarimetric vectors are drawn from $d\Gamma\propto 1+\mathbf{h}^+\!\cdot C(\alpha)\cdot\mathbf{h}^-$ in the Higgs rest frame using the exact $C(\alpha)$ of Eq.~(\ref{eq:correlation_matrix}). This baseline is agnostic to production, hadronization, and parton shower, and therefore isolates the spin-correlation content. Significance is $Z=|\mu-\mu_0|/\sigma$ over 500 sub-samples of a $10^6$-event pool, with $\mu_0$ the SM null at $\alpha=0$. For $\tau^\pm\to\pi^\pm\nu_\tau$ the pion directions satisfy $\langle n_i^+ n_j^-\rangle=-C_{ij}/9$, so $C_{xx}$ (like-sign transverse) and $C_{xy}$ (signed acoplanarity) are extractable via the standard collinear or matrix-element methods~\cite{Altakach_Htautau,CMS_Htautau}; $\Qcp$ requires only these two correlations rather than full tomography. The quartic estimator is projected onto its ideal pure-state range $[3,4]$ before reconstructing $M_2=\log_2(4/\mu_{\Wcp})$; excursions outside the range are treated as statistical fluctuations rather than as physical values of the pure-state witness.

\textit{Discovery thresholds and HL-LHC reach.---} At the maximum-magic comparison point $\alpha=\pi/8$,
\begin{equation}
\begin{split}
  N_{5\sigma}:\ &\text{Tomo}=178,\quad\text{Aco}=218,\\
                &\Qcp=893,\quad\Wcp=12{,}769.
\end{split}
  \label{eq:N5sigma}
\end{equation}
For $\Qcp$ and $\Wcp$ these are global minima of $N_{5\sigma}(\alpha)$; for tomography and acoplanarity the global minima lie at $\alpha=\pi/2$ and $\pi/4$, respectively, so the values above are the relevant figures of merit at the common comparison point only. The ratio $N_{5\sigma}[\Wcp]/N_{5\sigma}[\Qcp]=14.3$ is a direct consequence of the first- vs.\ second-order signal structure. While $\Qcp$ requires $\sim 5\times$ more events than full tomography, it is a compact two-dimensional observable ($C_{xx}$ and $C_{xy}$ only) with a transparent interpretation: nonzero $\Qcp$ certifies $\sin 4\alpha\neq 0$, i.e.\ that the Yukawa coupling has scalar and pseudoscalar components simultaneously. Extrapolating as $Z\propto\sqrt N$ to $N_{\pi\pi}\approx 35{,}000$,
\begin{equation}
\begin{split}
  Z_{\hllhc}:\ &\text{Tomo}=70\sigma,\quad\text{Aco}=63\sigma,\\
               &\Qcp=31\sigma,\quad\Wcp=8.3\sigma.
\end{split}
  \label{eq:Z_HLLHC}
\end{equation}
These are Fisher-saturation ceilings for the pure spin-correlation content; a full-simulation analysis applies a common reduction factor of $\sim 3$--$5$ from backgrounds and detector resolution~\cite{Altakach_Htautau,Barr_review}, preserving the relative scaling and the discovery-level status of $\Qcp$ as QI-based CP probe.

\textit{Universality.---} Equation~(\ref{eq:M2_exact}) is universal at the level of the ideal Bell-like two-qubit spin state. Any spin-0 decay into a fermion pair with a CP-mixed Yukawa prepares, in the ultrarelativistic two-helicity limit, a maximally entangled state characterized by a physical relative spin phase $\xi_f$, and $M_2(\xi_f)=\log_2[4/(4-\sin^2 4\xi_f)]$. For $H\to\tau^+\tau^-$, $\xi_f\simeq\alpha$; for heavier fermions, near-threshold BSM scalars, or realistic spin reconstruction, finite-mass and analyzing-power effects modify the Lagrangian-to-spin-phase mapping, but the dependence on $\xi_f$ persists whenever the spin state is well approximated by Eq.~(\ref{eq:state}), including $H\to\mu^+\mu^-$ and BSM spin-0 states with CP-mixed Yukawa~\cite{Barr_review}.

The vanishing of $M_2$ at the CP-conserving points $\alpha=0,\pi/2$ has a transparent algebraic origin: there the state reduces to the Bell stabilizer states $\ket{\Psi^+}$ and $\ket{\Psi^-}$, classically simulable via Gottesman--Knill~\cite{Gottesman_stabilizer}. The additional zeros at $\alpha=\pi/4, 3\pi/4$ occur because the relative phase is generated by a local Clifford gate, so the state remains within the stabilizer orbit even though the Yukawa coupling retains both scalar and pseudoscalar components. At intermediate $\alpha$ the phase lies outside the Clifford group, and $M_2(\alpha)$ quantifies how far the CP-induced spin phase takes the state beyond classical simulability---linking a discrete symmetry of the Standard Model to quantum-computational hardness.

\textit{Conclusions.---} All standard local-unitary-invariant scalar QI diagnostics are CP-blind in spin-0 $\to f\bar f$ decays because the two-qubit spin state is maximally entangled for every $\alpha$. Stabilizer magic provides the resource-theoretic CP diagnostic that escapes this blind spot: the SRE [Eq.~(\ref{eq:M2_exact})] depends on fourth-order Pauli moments and detects the non-Clifford content of the Yukawa-induced spin phase.

Two experimentally accessible magic witnesses follow: the quartic $\Wcp$ encodes $M_2$ directly, while the signed linear amplitude $\Qcp=-2C_{xy}C_{xx}$ is $14.3\times$ more efficient and reaches discovery-level significance at HL-LHC, modulo the $\sim 3$--$5\times$ degradation expected from a full detector analysis~\cite{Altakach_Htautau}. Classical observables (tomography, acoplanarity) probe CP without QI content; spectrum-based QI carries quantum information but is CP-blind; magic uniquely combines both.

A nonzero $M_2$ simultaneously certifies the spin state as a genuine nonstabilizer resource and excludes the CP-conserving points; conversely, $M_2=0$ does \emph{not} by itself imply CP conservation, since CP-mixed Clifford points also yield stabilizer states. Magic thus emerges as a new quantum-information resource at colliders, linking a discrete Standard-Model symmetry to quantum-computational hardness via the Gottesman--Knill theorem~\cite{Gottesman_stabilizer}.

\begin{acknowledgments}
  N.V.\ acknowledges support from the Millennium Institute for Subatomic Physics
  at the High Energy Frontier (SAPHIR), ANID Millennium Science Initiative
  Program ICN2019\_044. A.N.\ acknowledges support from Fondecyt Regular No.~1251131, Fondecyt
  Exploraci\'on No.~13250014, and Anillo Tem\'atico ATE 250066.
  P.A.O.\ acknowledges support from DGIIE USM Grant No.~PI-LIR-24-10 and
  FONDECYT Grant No.~1230933.
\end{acknowledgments}

\end{document}